\documentclass[12pt]{article}
\usepackage{graphicx}
\usepackage{amssymb}
\usepackage{amsmath}
\usepackage{esint}
\usepackage{dsfont}
\usepackage{cite}
\numberwithin{equation}{section}

\newenvironment{dedication}
        {\vspace{1ex}\begin{quotation}\begin{center}\begin{em}}    
        {\par\end{em}\end{center}\end{quotation}}
        
 \newsymbol \blackbox 1004
\newcommand{\eh}{\hfill}\newlength{\sperr}

\newenvironment{proof}{{\settowidth{\sperr}{\bf\rm
Proof}%
\par\addvspace{0.3cm}\noindent\parbox[t]{1.3\sperr}
{\bf\rm P\eh r\eh o\eh o\eh f\eh }%
}}{\nopagebreak\mbox{}
$\blackbox$\par\addvspace{0.3cm}}

\def\BC{{\mathbb C}}
\def\BR{{\mathbb R}}

\def\cll{{\mathcal L}}

\def\b{\beta}
\def\la{\lambda}
\def\om{\omega}
\def\ve{\varepsilon}
\def\ov{\overline}
\def\nn{\nonumber}
\def\wt{\widetilde}
\newtheorem{Pa}{Paper}[section]
\newtheorem{Tm}[Pa]{{\bf Theorem}}

\newtheorem{Cy}[Pa]{{\bf Corollary}}
\newtheorem{Rk}[Pa]{{\bf Remark}}

\newtheorem{Dn}[Pa]{{\bf Definition}}
\newtheorem{Pn}[Pa]{{\bf Proposition}}

\newcommand{\E}{\mathrm{e}}
\newcommand{\I}{\mathrm{i}}

\title{Dirac equation: the stationary and dynamical scattering problems}

\author{Lev Sakhnovich}
\date{}

\begin{document}
\maketitle

\begin{dedication}
Dedicated to Heinz Langer on the occasion of his eightieth
birthday with admiration.  
\end{dedication}

\begin{abstract}  We prove that for the  radial Dirac  equation
 with Coulomb-type
potential the generalized dynamical scattering operator
coincides with the corresponding generalized stationary
scattering operator. This fact is a quantum mechanical analogue of ergodic results
in the classical mechanics.  
\end{abstract}



\section{Introduction}\label{secIntro}
 In the present paper we consider radial Dirac systems with Coulomb-type potentials:
\begin{align} &
\left(\frac{d}{dr}+\frac{k}{r}\right)f-(\lambda+m-v(r))g=0,\label{2.1}
\\ &
\left(\frac{d}{dr}-\frac{k}{r}\right)g+(\lambda-m-v(r))f=0, \quad \, k=\overline{k}{\ne}0, \quad m>0.
\label{2.2}
\end{align}
We assume that the potential $v(r)$ has the form
\begin{equation}v(r)=-\frac{A}{r}+q(r),\quad A=\overline{A}{\ne}0,\quad |k|>|A|.\label{2.3}\end{equation}
We use the notions of the generalized wave operators, deviation factors and the generalized (dynamical) scattering operators $S_{dyn}$ (see \cite{Sakh2}). In  Section \ref{sec3} we introduce
the notions of the generalized stationary scattering operators $S_{st}$  and the corresponding deviation factors. The main result of this  paper is the following ergodic type equality 
(see Theorem \ref{Theorem 5.1}):
\begin{equation}S_{dyn}=S_{st}.\label{1.1}\end{equation}
Equality \eqref{1.1} is new even in the case $A=0$ (in \eqref{2.3}), which is treated separately  in Section \ref{secA}.
\begin{Rk}\label{Remark 1.1} The ergodic theorems in  classical mechanics
assert that, under certain conditions, the time average of a function along the trajectories exists almost everywhere and is related to the space average.
In quantum mechanics,
relation \eqref{1.1} is an analogue of the formulated ergodic properties from classical mechanics.\end{Rk}
\section{Radial Dirac system}\label{secDir}
In this section, we study the asymptotic behavior of the solutions of radial Dirac system \eqref{2.1}--\eqref{2.3}.
Introduce the following notations
\begin{equation} \gamma=\sqrt{k^{2}-A^{2}}>0,\quad \ve=\sqrt{\lambda^{2}-m^2}>0 \quad (|\lambda|>m),
\label{2.5}\end{equation}
where $\I$ is the imaginary unit. We deal with the two cases:
\begin{equation}\lambda>m,\quad \sqrt{m+\lambda}>0,\quad -\I\sqrt{m-\lambda}>0,\label{2.16}
\end{equation}
and
\begin{equation}\lambda<-m,\quad \sqrt{m-\lambda}>0,\quad -\I \sqrt{m+\lambda}>0.
\label{2.17}
\end{equation}
Further, $m>0$ is fixed and the formulas  below are valid (if not stated otherwise) for both cases.
We consider solutions of \eqref{2.1}, \eqref{2.2} depending on $r$, $k$ and $\lambda$  or on $r$, $k$ and $\ve$.
It is easy to see that in both cases \eqref{2.16} and \eqref{2.17} the variable $\lambda$ is uniquely recovered from
$\ve$.

1. We begin with the case when
\begin{equation}v(r)=-\frac{A}{r}\quad (A=\overline{A}{\ne}0,\quad |k|>|A|), \quad {\mathrm{i.e.,}} \quad
q(r)\equiv 0 ,
\label{2.4}\end{equation} 
The regular at the point $r=0$ solution $F_0=\begin{bmatrix}f_0 \\ g_0\end{bmatrix}$ of system \eqref{2.1}, \eqref{2.2}, with $v$ of the form \eqref{2.4}, satisfies the condition
\begin{equation}F_0=\begin{bmatrix}f_0 \\ g_0\end{bmatrix}{\sim}Nr^{\gamma}\begin{bmatrix}1 \\ b_0\end{bmatrix}, \quad
b_0:=(\gamma+k)/A,\quad r{\to}0,\label{2.6}\end{equation}
where $N$ does not depend on $r$ and $N{\ne}0$. \emph{Further we assume, that
$N=1$.}

The solution $F_0$ can be represented in the form (see \cite[Section 36]{BLP}):
\begin{equation}f_0=\sqrt{m+\lambda}\, \E^{-\I \ve r}r^{\gamma}(Q_{1}+Q_{2}),\quad
g_0=-\sqrt{m-\lambda}\, \E^{-\I \ve r}r^{\gamma}(Q_{1}-Q_{2}). \label{2.7}\end{equation}
The functions $Q_{1}$ and $Q_{2}$ can be expressed with the help of the confluent hypergeometric functions $\Phi(a,c,x)$ (see \cite{BaErd}):
\begin{align} & Q_1=a_1\Phi(\gamma-A\lambda/\ve, \, 2\gamma+1,\, 2\I \ve r),\label{2.8}
\\ & 
Q_2=a_2\Phi(\gamma+1-A\lambda/\ve,\, 2\gamma+1, \, 2\I \ve r).\label{2.9}\end{align}
Using relation \eqref{2.6} and equalities $N=1,\, \Phi(a,c,0)=1$ we have
\begin{equation}(a_2+a_1)\sqrt{m+\lambda}=\frac{A}{\gamma+k}(a_2-a_1)\sqrt{m-\lambda}=1.
\label{2.10}\end{equation}
System \eqref{2.1}, \eqref{2.2}  (where $v$  is given by  \eqref{2.4})
admits also a non-regular at $r=0$
solution $G_{0}=\begin{bmatrix}\phi_0 \\ \psi_0 \end{bmatrix}$
 of the form
\cite{AG, BaErd}:
\begin{equation}\phi_0=\sqrt{m+\lambda}\, \E^{-\I \ve r}r^{\gamma}(P_{1}+P_{2}),\quad
\psi_0=-\sqrt{m-\lambda}\, \E^{-\I \ve r}r^{\gamma}(P_{1}-P_{2}),\label{2.11}\end{equation}
where
\begin{align}& P_{1}=b_{1}\Psi(\gamma+A\I \lambda/\ve,\, 2\gamma+1,\, 2\I \ve r)
,\label{2.12}
\\ &
P_{2}=b_{2}\Psi(\gamma+1+A\I \lambda/\ve,\, 2\gamma+1,\, 2\I \ve r),
\label{2.13}\end{align}
and  $\Psi(a,c,x)$ is the confluent hypergeometric function of the second kind.
When $r{\to}0$, we have (see \cite[Ch. 6]{BLP}):
\begin{equation}\phi_0{\sim}\sqrt{m+\lambda}\, r^{-\gamma}(b_1+b_2),\quad
\psi_0{\sim}-\sqrt{m-\lambda}\, r^{-\gamma}(b_1-b_2),
\label{2.14}\end{equation}
where
\begin{equation}(b_2+b_1)\sqrt{m+\lambda}=\frac{A}{-\gamma+k}(b_2-b_1)\sqrt{m-\lambda}.
\label{2.15}\end{equation}
Using formulas \eqref{2.7}--\eqref{2.9}
and asymptotic behavior of the   confluent hypergeometric function
(see \cite{BLP}) we obtain the relations
\begin{align} &
f_0=2\Re\big(a_{1}\sqrt{m+\lambda}\, \E^{-\I \ve r}r ^{-\I A\lambda/{2\ve}}
C_{0}(-\ve)\big(1+M_{f}/r+O(r^{-2})\big)\big),\label{2.18}
\\ &
g_0=2\Re\big(a_{2}\sqrt{m-\lambda}\, \E^{\I \ve r}r ^{\I A\lambda/{2\ve}}
C_{0}(\ve)\big(1+M_{g}/r+O(r^{-2})\big)\big),\label{2.19}\end{align} 
where $M_f$ and $M_g$ do not depend on r, $\, r{\to}\infty,\, $  and
\begin{equation}C_{0}(\ve)=\frac{\Gamma(2\gamma+1)}
{\Gamma(\gamma+\I A\lambda/\ve)}(2\I \ve)^{-\gamma+\I A\lambda/(2\ve)}.\label{2.20}\end{equation}
Taking into account
\eqref{2.11}--\eqref{2.13} and asymptotic behavior of the   confluent hypergeometric function
of the second kind (see \cite{BLP}), we obtain the relations:
\begin{align}& \phi_0=b_{1}\sqrt{m+\lambda}\, \E^{-\I \ve r}(2 \I 
\ve)^{-\gamma}{(2\I \ve r)}^{-\I A\lambda/{\ve}}\big(1+M_{\phi}/r+O(r^{-2})\big),
\label{2.21}
\\ &
\psi_0=-b_{1}\sqrt{m-\lambda}\, \E^{-\I \ve r}(2\I \ve)^{-\gamma}{(2\I \ve r)}^{-\I A\lambda/{\ve}}\big(1+M_{\psi}/r+O(r^{-2})\big),
\label{2.22}
\end{align}
where $r{\to}\infty$ and $M_{\phi}$, $\, M_{\psi}$ do not depend on $r$.

2. \emph{Now, we consider the case $q(r){\not\equiv}0.$} That is, we consider system \eqref{2.1}, \eqref{2.2}, where
the initial  $v$ of the form \eqref{2.4} is perturbed by $q$ and has the  form \eqref{2.3}. We study this case using
solutions $F_0=\begin{bmatrix}f_0 \\ g_0 \end{bmatrix}$ and $G_0=\begin{bmatrix}\phi_0 \\ \psi_0 \end{bmatrix}$ of system  \eqref{2.1}, \eqref{2.2}, \eqref{2.4} and assume that
\begin{equation}\int_{0}^{\infty} (1+r)|q(r)|dr<\infty,\quad q(r)=\overline{q(r)}.\label{2.23}\end{equation}
Introduce  the $2{\times}2$ matrices $D(r,k,\ve)$ and $H(r)$  by the relations
\begin{equation}D(r,k,\ve)=\left[  \begin{array}{cc}
                         f_0(r,k,\ve) & \phi_0(r,k,\ve) \\
                         g_0(r,k,\ve) & \psi_0(r,k,\ve) \\
                         \end{array}
                       \right],\quad
                       H(r)=\left[\begin{array}{cc}
                             0 & q(r) \\
                             -q(r) & 0
                           \end{array}\right].\label{2.25}\end{equation}
It is easy to see, that
the  solution  $F(r,k,\ve)$ of
the integral equation
\begin{equation}F(r,k,\ve)=F_{0}(r,k,\ve)-\int_{0}^{r}D(r,k,\ve)D(t,k,\ve)^{-1}H(t)
F(t,k,\ve)dt \label{2.24}\end{equation} 
satisfies the  system \eqref{2.1}--\eqref{2.3} where
\eqref{2.23} holds.
\begin{Pn}\label{Proposition 2.1}The solution $F(r,k,\ve)$ of  \eqref{2.24} has the asymptotics
\begin{equation}F(r,k,\ve){\sim}{r}^{\gamma}\begin{bmatrix}1 \\ b_0\end{bmatrix},\quad 
r{\to}0.\label{2.26}\end{equation}\end{Pn}
\begin{proof}.  In view of \eqref{2.1}, \eqref{2.2}, $\det{D(r)}$ does not depend on $r$.  Hence, using  \eqref{2.6} and \eqref{2.14}   we obtain
\begin{equation}\det{D(r,k,\ve)}=M_d(k,\ve){\ne}0.\label{2.27}\end{equation}
Thus, we have
\begin{equation}r^{-\gamma}D(r)D^{-1}(t)t^{\gamma}=O(1),\quad r{\geq}t,\quad r{\to}0.\label{2.28}\end{equation}
The proposition follows from \eqref{2.23}, \eqref{2.24} and \eqref{2.28}.
\end{proof}
Consider the solution $\Phi_0$ of \eqref{2.1}, \eqref{2.2}, \eqref{2.4} with the  asymptotics
\begin{equation}\Phi_0(r,k,\ve)=\E^{-\I \ve r}r^{\I A\lambda/\ve}
\begin{bmatrix}\sqrt{m+\lambda}+o(1) \\ -\sqrt{m-\lambda}+o(1) \end{bmatrix},
\quad r{\to}\infty.\label{2.30-}\end{equation}
The  solution  $\Phi(r,k,\ve)$ of
the integral equation
\begin{equation}\Phi(r,k,\ve)=\Phi_{0}(r,k,\ve)+\int_{r}^{\infty}D(r,k,\ve)D(t,k,\ve)^{-1}H(t)
\Phi(t,k,\ve)dt\label{2.29}\end{equation} 
satisfies the system \eqref{2.1}--\eqref{2.3}.
\begin{Pn}\label{Proposition 2.2}The solution $\Phi(r,k,\ve)$ of  \eqref{2.29} has the asymptotics
\begin{equation}\Phi(r,k,\ve)=\E^{-\I \ve r}r^{\I A\lambda/\ve}
\begin{bmatrix}\sqrt{m+\lambda}+o(1) \\ -\sqrt{m-\lambda}+o(1) \end{bmatrix},
\quad r{\to}\infty.\label{2.30}\end{equation}
\end{Pn}
\begin{proof}. Taking into account  \eqref{2.18}, \eqref{2.19} and \eqref{2.21}, \eqref{2.22}, we derive
the relation
\begin{equation}||D(r)D^{-1}(t)]||=O(1)\quad (r{\leq}t),\quad r{\to}\infty.
\label{2.31}\end{equation}
The proposition follows from \eqref{2.29}  and \eqref{2.31}.
\end{proof}
Let us consider  $\Phi(r,k,\ve)=\begin{bmatrix}\phi(r,k,\ve) \\ \psi(r,k,\ve) \end{bmatrix}$ in greater detail.
Using \eqref{2.21}--\eqref{2.23} and \eqref{2.29}, we obtain the assertion below.
\begin{Cy}\label{Corollary 2.3}The entries of the solution $\Phi(r,k,\ve)$ of  \eqref{2.29} have the following asymptotics:
\begin{align}& \phi=b_{1}\sqrt{m+\lambda}\, \E^{-\ve r}(2\ve)^{-\gamma}{(2\ve r)}^{A\lambda/{\ve}}\big(1+M_{\phi}/r+O(m(r))\big),
\label{2.32}
\\ &
\psi=-b_{1}\sqrt{m-\lambda}\, \E^{-\ve r}(2\ve)^{-\gamma}{(2\ve r)}^{A\lambda/{\ve}}\big(1+M_{\psi}/r+O(n(r))\big),
\label{2.33}
\end{align}
where $r\to \infty$, $M_{\phi}$ and $M_{\psi}$ do not depend on $r$, 
and the inequality
\begin{equation}\int_{a}^{\infty}(|m(r)|+|n(r)|)dr<\infty \label{2.34}\end{equation}\end{Cy}
is valid for some $a>0$.
\section{Generalized stationary scattering operators}\label{sec3}
 In many important cases, the initial and final states of the system (i.e., the states when $t{\to}\pm\infty$) cannot be regarded as free. 
 For these cases the generalized dynamical  wave operators  and generalized
dynamical scattering operators are used effectively instead of the usual dynamical  wave and scattering operators (see \cite{BuM, Sakh8, Sakh2}). In the present section we  consider the radial Dirac system 
\eqref{2.1}, \eqref{2.2} and introduce  the notions of generalized wave and scattering operators for the stationary case. In this way, we deal with the non-free states when $r\to \pm \infty$
(instead of the non-free states when  $t{\to}\pm\infty$ for dynamical systems).
Let  the following condition
\begin{equation}\int_{a}^{\infty}|q^{\prime}(r)|dr+
\int_{a}^{\infty}|q^{2}(r)|dr+ \int_{b}^{a}|q(r)|dr<\infty,\quad 0<b<a<\infty, \label{3.1}\end{equation}
where $q^{\prime}(r):= \big(\frac{d}{dr}q\big)(r)$, be fulfilled.
System \eqref{2.1}, \eqref{2.2} can be written in the matrix form
\begin{equation}\frac{d}{dr}Z=\mathcal{A}(r)Z,\quad \mathcal{A}(r)=\left[\begin{array}{cc}
                  -k/r & m+\lambda-v(r)\\
               m-\lambda+v(r)    & k/r
                \end{array}\right], \label{3.2}\end{equation}
where $Z(r,k, \la)\in \BC^2$. According to  \cite[Ch.II, Theorem 8]{Bel}, 
system \eqref{3.2} has two linear independent solutions $Z_1$ and $Z_2$ such that:
\begin{align}& Z_{1}(r,k,\la){\sim}\exp\{-\I {\theta}\}V_{0}(r,\lambda)^{-1}C_{1}(k,\lambda),\quad r{\to}\infty;  \label{3.4}
\\ &
Z_{2}(r,k,\lambda){\sim}\exp\{\I{\theta}\}V_{0}(r,\lambda)C_{2}(k,\lambda),\quad r{\to}\infty,  \label{3.5}\end{align}
where ${\theta}={\ve}r$, 
 $C_1(k,\lambda)$ and $C_2(k,\lambda)$ are $2{\times}1$ vectors, $C_1(k,\lambda)=\overline{C_2(k,\lambda)}$, $\ve$ is introduced in \eqref{2.5}, and
\begin{equation}V_{0}(r,\lambda)=\exp\left\{\I \frac{\lambda}{\ve}\int_{a}^{r}v(u)du\right\}. \label{3.6}\end{equation}
Recall that we consider the cases \eqref{2.16} and \eqref{2.17}, and (if not stated otherwise) our formulas  are valid  for both cases.  
Relations \eqref{3.4} and \eqref{3.5} yield   the next assertion.
\begin{Pn}\label{Proposition 3.1}Let condition \eqref{3.1} be fulfilled. Then, the regular at the point $r=0$ solution $Z_{reg}$ of system \eqref{2.1}--\eqref{2.3} has
the following asymptotics at $r\to \infty\, :$
\begin{equation}Z_{reg}{\sim}\frac{1}{2\I }\big(\exp\{\I {\theta}\}V_{0}(r,\lambda)C_2(k,\lambda)-
\exp\{-\I{\theta}\}V_{0}(r,\lambda)^{-1}C_ 1(k,\lambda)\big).
\label{3.7}\end{equation}
\end{Pn}
We introduce the scattering matrix function via the entries of $$C_1(k,\lambda)=\begin{bmatrix}c_{1,1}(k,\lambda) \\ c_{2,1}(k,\lambda)\end{bmatrix}.$$
\begin{Dn}\label{Definition 3.2} The matrix function 
\begin{equation}S(k,\lambda)=\left[
\begin{array}{cc}
   s_{1,1}(k,\lambda) & 0 \\
   0 & s_{2,1}(k,\lambda)
 \end{array}\right],
 \label{3.9}\end{equation}
 where
\begin{equation} 
  s_{n,1}(k,\lambda):=c_{n,1}(k,\lambda)\big/ \, \overline{c_{n,1}(k,\lambda)} \quad (n=1,2),
 \label{3.8} \end{equation} 
is called the generalized stationary   scattering matrix function.\end{Dn}
\begin{Dn}\label{Definition 3.3}The function $V_{0}(r,\lambda)$ $($see \eqref{3.6}$)$
is called the stationary deviation factor.\end{Dn}
It follows from \eqref{3.9}, \eqref{3.8} and the equality $C_1(k,\lambda)=\overline{C_2(k,\lambda)}$  that
\begin{equation}S(k, \lambda)C_2(k,\lambda)=C_1(k,\lambda).\label{3.10}
\end{equation}
\begin{Rk}\label{Remark 3.4} Note that the deviation factor $V_{0}(r,\lambda)$ does not depend on $k$.
\end{Rk}

\section{Coulomb-type potentials: spectral theory}\label{secCoul}
1. We study first the  system \eqref{2.1}, \eqref{2.2} where
\begin{equation} k=0, \quad v(r)\equiv 0,
\label{4.0}\end{equation}
that is, $k=0$, $A=0$, $q(r)\equiv 0$. In this case we have the system:
\begin{align} &\frac{d}{dr}f-(\lambda+m)g=0,
\label{4.1}
\\ &
\frac{d}{dr}g+(\lambda-m)f=0, \quad m>0,\quad 0{\leq}r<\infty .\label{4.2}
\end{align}
Consider the special solution $f_1=f$ and $g_1=g$ of \eqref{4.1}, \eqref{4.2} with the initial conditions
\begin{equation}f_1(0,\lambda)=1,\quad g_1(0,\lambda)=0.\label{4.3}\end{equation}
It follows from \eqref{4.1} and \eqref{4.2} that both $f(r,\lambda)$ and
$g(r,\lambda)$ satisfy the equation:
\begin{equation}\frac{d^2}{dr^2}y+(\lambda^2-m^2)y=0.\label{4.4}\end{equation}
Taking into account \eqref{4.3} and  \eqref{4.4}, we obtain the equalities
\begin{equation}f_{1}(r,\lambda)=\cos\ve{r},\quad g_{1}(r,\lambda)=\beta(\lambda)\sin\ve{r},
\label{4.5}\end{equation}
where $\ve$ is given in \eqref{2.5}. Recall that either \eqref{2.16} or \eqref{2.17} holds.
Relations \eqref{4.2} and  \eqref{4.3} imply that
\begin{equation}\frac{d}{dr}g_{1}\Big|_{r=0}=m-\lambda.\label{4.7}\end{equation}
It is immediate from \eqref{4.5} and \eqref{4.7} that
\begin{equation}\beta(\lambda)= (m-\lambda)/\ve. \label{4.8}\end{equation}

2. Let us introduce the  differential operator $\mathcal{L}_{0}$, which corresponds to Dirac system \eqref{4.1}, \eqref{4.2}:
\begin{equation}\big(\mathcal{L}_{0}h\big)(r)=j_{1}\frac{d}{dr}h(r)-mj_{2}h(r) \quad
 (0{\leq}r<\infty),\label{4.16}\end{equation}
where 
\begin{equation}j_{1}=\left[\begin{array}{cc}
                        0 & -1 \\
                        1 & 0
                      \end{array}\right],\quad
j_{2}=\left[\begin{array}{cc}
                        -1 & 0 \\
                        0 & 1
                      \end{array}\right],
\quad h(r)=\begin{bmatrix} h_1(r) \\ h_2(r) \end{bmatrix}.                      
                      \label{4.17}\end{equation}
The boundary condition is defined by the relation:
\begin{equation}h_{2}(0)=0.\label{4.18}\end{equation}
By $\mu_{1}(u)$, we denote  the spectral function of the operator $\mathcal{L}_{0}$
Using \eqref{4.5} and \eqref{4.8}, it is easy to see that the spectral density $\rho_{1}(\sigma)=\mu_{1}^{\prime}(\sigma)$  is given by the formula
\begin{equation}\rho_{1}(\sigma)=\frac{1}{\pi}\sqrt{\Big|\frac{\sigma+m}{\sigma-m}\Big|} \quad {\mathrm{for}} \quad |\sigma|>m; \quad \rho_{1}(\sigma)\equiv 0
\quad {\mathrm{for}} \quad |\sigma|<m.
\label{4.15}\end{equation}
The operator ${\mathcal{L}}_{0}$ is similar to the multiplication by $\lambda$ in a space of functions $F(\la)$
($\la \in E$) where
\begin{align}& \label{E}
E:=(-\infty, \, -m] \, {\bigcup} \, [m,\, +\infty).
\end{align} 
More precisely, we have the following proposition.
\begin{Pn}\label{Proposition 4.1}The operator $\mathcal{L}_{0}$, introduced by \eqref{4.16}, \eqref{4.18}, admits representation
\begin{equation}\mathcal{L}_{0}=U_{0}QU_{0}^{-1}, \label{4.19}\end{equation}
where the operators $U_{0}$, $U_{0}^{-1}$  and $Q$
are given by the equalities$\,:$
\begin{align} & \big(U_{0}F\big)(r)=\int_{E}\begin{bmatrix} f_1(r,\la) \\ g_1(r,\la) \end{bmatrix}F(\lambda)
\rho_{1}(\lambda)d\lambda=h(r),\label{4.20}
\\ &
\big(U_{0}^{-1}h\big)(\la)=\int_{0}^{\infty}\begin{bmatrix} f_1(r,\la) & g_1(r,\la) \end{bmatrix} h(r)dr=F(\lambda),
\label{4.21}
\\ &
\big(QF\big)(\lambda)=\lambda{F(\lambda)}\quad
(\lambda{\in}E).\label{4.22}\end{align}
\end{Pn}
\begin{proof}. In view of \eqref{4.16},  \eqref{4.20} and \eqref{4.22}, direct calculation shows that
\begin{equation}\mathcal{L}_{0}U_{0}F=U_{0}QF.\label{4.23}
\end{equation}
\end{proof}
We note that the following Parseval-type relation holds (see \cite[Ch. 10]{Sakh3}).
\begin{Pn}\label{Proposition 4.2} Let equality \eqref{4.20} be valid. Then, we have
\begin{equation}\int_{E}|F(\lambda)|^{2}\rho_{1}(\lambda)d\lambda=
\int_{0}^{\infty}\big(|h_{1}(r)|^{2}+|h_{2}(r)|^{2}\big)dr,\label{4.24}
\end{equation}
where $h_k$ $(k=1,2)$ are the entries of $h$.
\end{Pn}
3. Next, we consider  the radial Dirac system \eqref{2.1}--\eqref{2.3} assuming that 
\eqref{2.23} holds.  We introduce the differential operator
\begin{equation}\mathcal{L}h(r)=j_{1}\frac{d}{dr}h(r)-mj_{2}+V(r)h(r),\quad
 0{\leq}r<\infty,\label{4.25}\end{equation}
where
\begin{equation}V(r)=\left[\begin{array}{cc}
                             v(r) & k/r \\
                             k/r & v(r)
                           \end{array}\right],
\quad h(r)=\begin{bmatrix} h_1(r) \\ h_2(r) \end{bmatrix}.                           
                           \label{4.26}\end{equation}
Similar to \eqref{4.18}, the boundary condition for  $\mathcal{L}$ is given by the relation:
\begin{equation}h_{2}(0)=0.\label{4.27}\end{equation}
 By $G$ we
denote the maximal invariant subspace on which the operator $\mathcal{L}$ induces an
operator with absolutely continuous spectrum, and $P$ stands for the orthogonal projection
from $L^2_2(0,\infty)$ onto $G$. The  spectral function  of the operator $\mathcal{L}$
 (given by \eqref{4.25} and \eqref{4.27}) is denoted by $\mu(\la)$ and the spectral density $\mu^{\prime}(\lambda)$
 (on the absolutely continuous part of spectrum)
 is denoted by $\rho(\lambda)$.
\begin{Pn}\label{Proposition 4.3}Let the operator  $\mathcal{L}$ be given by  \eqref{4.25}--\eqref{4.27} and \eqref{2.3}.
Assume that \eqref{2.23} holds.
  Then,  the equality
\begin{equation}\mathcal{L}P=UQU^{-1}P, \label{4.28}\end{equation}
where the operators $U$, $U^{-1}$  and $Q$
have the form
\begin{align}& \big(UF\big)(r)=\int_{E}\begin{bmatrix} f(r,\la) \\ g(r,\la) \end{bmatrix}F(\lambda)
\rho(\lambda)d\lambda=h(r){\in}G,\label{4.29}
\\ &
\big(U^{-1}h\big)(\la)=\int_{0}^{\infty}\begin{bmatrix} f(r,\la) & g(r,\la) \end{bmatrix}h(r)dr=F(\lambda) \quad (h{\in}G),
\label{4.30}
\\ &
\big(QF\big)(\lambda)=\lambda{F(\lambda)},\quad
\lambda{\in}E, \quad E:=(-\infty,\, -m]\, {\bigcup}\, [m,\, +\infty),\label{4.31}
\end{align}
is valid.
\end{Pn}
\begin{proof}. Similar to the proof of Proposition \ref{Proposition 4.1},
direct calculation (using \eqref{4.25}, \eqref{4.29} and \eqref{4.31} shows that
\begin{equation}\mathcal{L}UF=UQF.\label{4.32}
\end{equation} 
\end{proof}
The following Parseval-type relation is fulfilled (see \cite[Ch. 10]{Sakh3}).
\begin{Pn}\label{Proposition 4.4} Let the conditions of Proposition \ref{Proposition 4.3} hold. Then,
\begin{equation}\int_{E}|F(\lambda)|^{2}\rho(\lambda)d\lambda=
\int_{0}^{\infty}\big(|h_{1}(r)|^{2}+|h_{2}(r)|^{2}\big)dr.\label{4.33}
\end{equation}
\end{Pn}
\begin{Rk} According to \eqref{4.19} and \eqref{4.28}, we have
\begin{equation}\E^{\I t\mathcal{L}_{0}}\E^{-\I t\mathcal{L}}P=
U_{0}\E^{\I tQ}U_{0}^{-1}U\E^{-\I tQ}U^{-1}P.\label{4.34}\end{equation}
\end{Rk}

4. Taking into account \eqref{2.30}, we see that (under condition \eqref{2.20} instead of condition \eqref{3.1} in Section \ref{sec3}) 
the relations \eqref{3.7}--\eqref{3.10} are valid for $V_0$ of the form
\begin{equation}V_{0}(r,\lambda)=r^{\I A\lambda/\ve}.\label{4.35}\end{equation}
Using \eqref{3.7}--\eqref{3.10}, we consider below the generalized stationary scattering matrix
$S_{st}(\cll,\cll_0)=S(k,\la)$ for the case of Coulomb-type potentials satisfying \eqref{2.20}.

In view of Corollary \ref{Corollary 2.3}, we rewrite \eqref{3.7} as
\begin{equation}
Z_{reg}= \frac{1}{2\I}\om(k,\ve)\big(\ov{\Phi(r,k,\ve)}-\Phi(r,k,\ve)\big),
\label{4.35'}
\end{equation}
where $\om(k,\ve)$ is a real-valued function. It follows from \eqref{2.32} and \eqref{2.33} that
\begin{equation}c_{2,1}(k,\lambda)\big/\, c_{1,1}(k,\lambda)=\I \beta(\lambda),\label{4.36}\end{equation}
where $\b$ coincides  with $\b$ in  \eqref{4.8}. Relations \eqref{3.8} and \eqref{4.36} imply that
\begin{equation}s_{1,1}(k,\lambda)=-s_{2,1}(k,\lambda).\label{4.37}\end{equation}
Hence, the scattering matrix $S(k,\lambda)\,$ ($\lambda=\overline{\lambda}$) has the form
\begin{equation}S_{st}(\cll,\cll_0):=S(k,\lambda)=\left[
\begin{array}{cc}
   s_{1,1}(k,\lambda) & 0 \\
   0 &-s_{1,1}(k,\lambda)
 \end{array}\right], \quad |\lambda|>m.\label{4.38}\end{equation}

\section{Ergodic properties}\label{ErgProp}
In the present section,  we consider the generalized dynamical scattering operators (which are introduced in Appendix)
for the case of Dirac systems with Coulomb-type potentials, where $A_0=\cll_0=\cll_0^*$ and $A=\cll=\cll^*$, that is, we consider 
$S(\cll, \cll_0)$ given by \eqref{1.14} and \eqref{4.16}, \eqref{4.25}.  Moreover, we consider $S(\cll, \cll_0)$ in momentum representation
\begin{align}\label{d1} &
S_{dyn}(\cll, \cll_0)=U_0^{-1}S(\cll, \cll_0)U_0,
\end{align}
where $U_0$ and $U_0^{-1}$ are given by \eqref{4.20} and \eqref{4.21}
We compare the generalized dynamical scattering operator $S_{dyn}$ with the generalized stationary scattering operator $S_{st}$ given by \eqref{4.38}.
More precisely, we compare the actions of $S_{dyn}$ and $S_{st}$ on the subspace $L$ of functions $f(\la)\in \BC^2$ ($\la \in \wt E$), where \\ $\wt E=(-\infty, -m)\cup (m,\infty)$:
\begin{align}\label{d0} &
L=\left\{f(\la)=\begin{bmatrix} f_1(\la) \\ f_2(\la) \end{bmatrix}: \, f_1(\la) \equiv 0 \,\, {\mathrm{for}} \,\, \la < -m, \,\, f_2(\la) \equiv 0 \,\, {\mathrm{for}} \,\, \la >m\right\}.
\end{align}
 In this way, we find formulas, which demonstrate quantum analogues of the ergodic properties from classical mechanics.

\begin{Tm}\label{Theorem 5.1} Let the radial Dirac system \eqref{2.1}-\eqref{2.3} and corresponding operators
$\cll_0$ and $\cll$ $($defined via \eqref{4.16}, \eqref{4.18} and  \eqref{4.25}, \eqref{4.27}, respectively$)$ be given.
Assume that  \eqref{2.23} holds. Then, the generalized stationary and dynamical scattering
matrices are equal on $L$, that is, 
\begin{align}\label{d2} &
S_{st}(\mathcal{L},\mathcal{L}_{0})f=S_{dyn}(\mathcal{L},\mathcal{L}_{0})f \quad {\mathrm{for}} \quad f\in L.
\end{align}
\end{Tm}
\begin{proof}. Step 1. First, we study the operator $T=U_{0}^{-1}U$, where $U_0^{-1}$ and $U$ are given by \eqref{4.21} and \eqref{4.29}.
 According to \eqref{4.21} and \eqref{4.29},  $T$    admits the representation
\begin{equation}\big(TF\big)(\lambda)=\int_{E}F(u)\rho(u)\int_{0}^{\infty}
\big(f_{1}(r,\lambda)f(r,u)+g_{1}(r,\lambda)g(r,u)\big)drdu.\label{5.1}\end{equation}
Using \eqref{4.5} and \eqref{5.1}, we rewrite the operator $T$ in the form 
\begin{align} & T=T_1+T_2,  \label{d3}
\end{align}
where  the operators $T_1$  and $T_2$ are defined by the formulas
\begin{align}&
\big(T_{1}F\big)(\lambda)=
\frac{d}{d\ve}\int_{E}F(u)T_{1}(\ve,u)du, 
\label{5.2}
\\ &
T_1(\ve,u):= \rho(u)\int_{0}^{\infty}
f(r,u)\frac{\sin(\ve r)}{r}dr;
\label{5.4}
\\  &
\big(T_{2}F\big)(\lambda)=\beta(\lambda)
\frac{d}{d\ve}\int_{E}F(u)T_{2}(\ve,u)du, 
\label{5.3}
\\ &
 T_2(\ve,u):=
\rho(u)\int_{0}^{\infty}
g(r,u)\frac{1-\cos(\ve r)}{r}dr.
\label{5.5}\end{align}
We note  that $\beta(\lambda)$ in \eqref{5.3} is given by  \eqref{4.8}.

We shall need  some properties of the operators
\begin{equation}R_{l}g=
\frac{d}{d\ve}\int_{m}^{\infty}g(u)R_{l}(\ve,u)du,\quad l=0,1,2,
\label{5.6}\end{equation}
where the kernels $R_{l}(\ve,u)$ have the form
\begin{equation}R_{l}(\ve,u)=
\rho(u)\int_{0}^{\infty}p_{l}(r,u)\frac{1-\cos{\ve r}}{r}dr.
\label{5.7}\end{equation}
Here, $\rho$ is again the spectral density of $\cll$, the functions $p_0$ and $p_1$ are fixed, and $p_2$ is some summable function:
\begin{align} & p_{0}(r,u)=\E^{-\I \ve r}r^{-\I Au\big/\sqrt{u^2-m^2}}, 
\label{5.8}
\\ &
p_{1}(r,u)=p_{0}(r,u)\big/r,\quad \int_{0}^{\infty}|p_{2}(r,\ve)|dr<\infty.
\label{5.9}\end{align}
We assume that the functions
$g(u)$
 belong to the class $S$, that is,
 $g \in C^{\infty}$ and functions $g$ have finite support
 (more precisely, $g(u)\equiv 0$
for  $u\, {\notin}\, (a_g,\, b_g),$ where $m<a_g<b_g<\infty$).
It follows from  \eqref{5.6}--\eqref{5.9}   that
\begin{equation}R_{l}\big(\E^{\I t\sqrt{u^2-m^2}}g(u)\big){\to}0,\quad     t{\to}\pm\infty,\quad l=1,2.\label{5.10}\end{equation}
 Now, consider $R_0$.
 Recall the relation \cite[Ch. 2]{GSH}:
\begin{equation}\int_{0}^{\infty}r^{\mu}\E^{\I xr}dr=\I \Gamma(\mu+1)
\big(\E^{\I \mu\pi/2}x_{+}^{-\mu-1}-\E^{-\I \mu\pi/2}x_{-}^{-\mu-1}\big) ,
\label{5.11}\end{equation} 
where $\mu{\ne}-1,-2,...$;
$\, \Gamma(\zeta)$ is Euler gamma function; $x_{+}=x$ if $x>0$ and $x_{+}=0$ if $x<0$,
$x_{-}=0$ if $x>0$ and $x_{-}=|x|$ if $x<0$.
Due to relations  \eqref{5.6}--\eqref{5.8}  the operator $R_0$
can be written in the form
\begin{equation}R_{0}g=\frac{1}{2}
\frac{d}{d\ve}\int_{m}^{\infty}g(u)\big(\Phi(u,-\ve-\sqrt{u^2-m^2})+\Phi(u,\ve-\sqrt{u^2-m^2})\big)du,
\label{5.12}\end{equation}
where
\begin{align}& \Phi(u,\zeta)=-\Gamma\big(-\I \phi(u)\big)\rho(u)
\big(\E^{\pi \phi(u)/2}\zeta_{+}^{\I \phi(u)}+\E^{-{\pi \phi(u)/2}}\zeta_{-}^{\I {\phi(u)}}\big),
\label{5.13}
\\ &
\phi(u)=\frac{Au}{\sqrt{u^2-m^2}}.\label{5.14}
\end{align}
Introduce  the operators
\begin{equation}R_{\pm}g=
\frac{1}{2}\frac{d}{d\ve}\int_{m}^{\infty}g(u)\Phi(u,-\sqrt{u^2-m^2}{\mp}\ve)du,
\label{5.15}\end{equation}
According to \eqref{5.15}, we have
\begin{equation}R_{0}=R_{-}+R_{+}.\label{5.16}\end{equation}
It is easy to see that
\begin{equation}R_{+}\big(\E^{\I t\sqrt{u^2-m^2}}g(u)\big){\to}0,\quad
t{\to}\pm\infty.\label{5.17}\end{equation}

Step 2. In this and the following steps of proof, we consider  the case \eqref{2.16}, where $\la >m$. The case \eqref{2.17}
may be considered in the same way. The present step of proof is dedicated to the study of $R_-$.
Taking into account \eqref{5.13} and \eqref{5.15} we obtain
\begin{align}& R_{-}=V_{1}+V_{2},\label{5.18}
\\ &
V_{1}g=\frac{1}{2}\frac{d}{d\ve}\int_{m}^{\sqrt{m^2+\ve^2}}\, g(u)\xi_{1}(u)(\ve-\sqrt{u^2-m^2})^{\I\phi(u)}du,
\label{5.19}
\\ &
V_{2}g=\frac{1}{2}
\frac{d}{d\ve}\int_{\sqrt{m^2+\ve^2}}^{\infty}g(u)\xi_{2}(u)(\sqrt{u^2-m^2}-\ve)^{\I \phi(u)}du,
\label{5.20}\end{align}
where the functions $\xi_{1}(u)$ and $\xi_{2}(u)$ are defined by the relations
\begin{align}& \xi_{1}(u)=-\Gamma\big(-\I \phi(u)\big)\rho(u)\E^{\pi \phi(u)/2},  \label{5.21}
\\ &
\xi_{2}(u)=-\Gamma\big(-\I \phi(u)\big)\rho(u)\E^{-\pi \phi(u)/2}.\label{5.22}
\end{align}
The operator $V_{1}$  can be written in the form
\begin{align}\nn V_{1}g=\frac{1}{2}\frac{d}{d\ve}\int_{0}^{\ve} &
g\big(\sqrt{m^2+\eta^2}\big)\xi_{1}\big(\sqrt{m^2+\eta^2}\big)
\\ & \times
\big(\ve-\eta\big)^{\I \phi\big(\sqrt{m^2+\eta^2}\big)}\big(\eta/\sqrt{m^2+\eta^2}\big)d\eta.
\label{5.23}
\end{align}
 Using \cite[f-las (3.14) and (3.15)]{Sakh8},  we obtain (for  $t{\to}{\pm}\infty$):
\begin{equation}V_{1}\big(\E^{\I t\sqrt{u^2-m^2}}g(u)\big){\sim}\frac{\ve}{2\lambda}\E^{\I t\ve}|t|^{-\I \phi(\lambda)}
\Gamma\big(1+\I \phi(\lambda)\big)\E^{\pm\pi \phi(\lambda)/2}\xi_{1}(\lambda)g(\lambda), \label{5.24}\end{equation}
where $\lambda>m$ and the functions $g(u)$ belong to the class $S$.
Recalling the well-known relation
\begin{equation}\Gamma\big(1+\I \phi(\lambda)\big)\Gamma\big(-\I\phi(\lambda)\big)= \frac{\I \pi}{\sinh(\pi{\phi(\lambda)})},\label{5.25}\end{equation}
from  \eqref{5.21} and  \eqref{5.24} we derive  (for  $t{\to}{\pm}\infty$):
\begin{equation}|t|^{\I\phi(\lambda)}\E^{-\I t\ve}V_{1}\big(\E^{\I t\sqrt{u^2-m^2}}g(u)\big){\sim}
\nu_{1}(\lambda)\E^{\pm\pi\phi(\lambda)/2}g(\lambda),
\label{5.26}\end{equation}
where
\begin{equation}\nu_{1}(\lambda)=- \I \rho(\lambda)\frac{\pi\ve}{2{\lambda}\sinh(\pi\phi(\lambda))}\E^{\pi\phi(\lambda)/2}.
\label{5.27}\end{equation}
The operator $V_{2}$ given by \eqref{5.20} admits representation
\begin{align}\nn V_{2}g=\frac{1}{2}
\frac{d}{d\ve}\int_{\ve}^{\infty}& g\big(\sqrt{\eta^2+m^2}\big)\xi_{2}\big(\sqrt{\eta^2+m^2}\big)
\\ & \times
\big(\eta-\ve\big)^{\I\phi\big(\sqrt{\eta^2+m^2}\big)}
\big(\eta/\sqrt{\eta^2+m^2}\big)d\eta.
\label{5.28}\end{align}
Due to \eqref{5.28} we have
\begin{equation}V_{2}\big(\E^{\I t\sqrt{u^2-m^2}}g(u)\big){\sim}\lim_{\delta{\to}+0}
\frac{\ve}{2\I \lambda}\phi(\lambda)g(\lambda)\xi_{2}(\lambda)\int_{\ve}^{\infty}\E^{\I t \eta}\big(\eta-\ve\big)^{\I \phi(\lambda)+\delta-1}d\eta .
\label{5.29}
\end{equation}
Using  \eqref{5.11} and the equality
\begin{equation}\int_{\ve}^{\infty}\E^{\I t \eta}\big(\eta-\ve\big)^{\I \phi(\lambda)+\delta-1}d\eta=
\E^{\I t\ve} \int_{0}^{\infty}\E^{\I tr}r^{\I \phi(\lambda)+\delta-1}dr ,\label{5.30}\end{equation}
we obtain
\begin{align}\nn
\int_{\ve}^{\infty}\E^{\I t \eta}\big(\eta-\ve\big)^{\I\phi(\lambda)+\delta-1}d \eta {\to} & \E^{\I t\ve}
\Gamma\big(\I \phi(\lambda)\big)
\\  & \times
\big(\E^{-\pi\phi(\lambda)/2}t_{+}^{-\I \phi(\lambda)}+\E^{\pi\phi(\lambda)/2}t_{-}^{-\I \phi(\lambda)}\big),
\label{5.31}
\end{align}
where $\delta{\to}+0$.
It follows from \eqref{5.29} and \eqref{5.31} that for $t{\to}\pm\infty$ we have:
\begin{equation}V_{2}\big(\E^{\I t\sqrt{u^2-m^2}}g(u)\big){\sim}-\frac{\ve}{2\lambda}\E^{\I t\ve}|t|^{-\I \phi(\lambda)}g(\lambda)\xi_{2}(\lambda)
\Gamma\big(1+\I\phi(\lambda)\big)\E^{\mp\pi \phi(\lambda)/2}.\label{5.32}\end{equation}
Relations \eqref{5.22} and \eqref{5.32} imply that
\begin{equation}|t|^{\I \phi(\lambda)}\E^{-\I t\ve}V_{2}\big(\E^{\I t\sqrt{u^2-m^2}}g(u)\big){\sim}
\nu_{2}(\lambda)\E^{\mp \pi\phi(\lambda)/2}g(\lambda)\quad (t{\to}{\pm}\infty),
\label{5.33}\end{equation}
where
\begin{equation}\nu_{2}(\lambda)=\I \rho(\lambda) \frac{\pi\ve}{{2\lambda}\sinh\big(\pi \phi(\lambda)\big)}\E^{-\pi\phi(\lambda)/2}.
\label{5.34}\end{equation}
Finally, taking into account  \eqref{5.16}--\eqref{5.18}, \eqref{5.26} and \eqref{5.33}
we have
\begin{equation}R_{0}\big(\E^{\I t\sqrt{u^2-m^2}}g\big){\sim} \frac{\pi\ve}{\I \lambda}|t|^{-\I \phi(\lambda)}\E^{\I t\ve}\rho(\lambda)g(\lambda),
\quad t{\to}+\infty,
\label{5.35}\end{equation}
and
\begin{equation}R_{0}\big(\E^{\I t\sqrt{u^2-m^2}}g(u)\big){\sim}0,
\quad t{\to}-\infty.
\label{5.36}\end{equation}
It is easy to see that the equality 
$$\overline{R}_{0}\big(\E^{\I t\sqrt{u^2-m^2}}g(u)\big)=\overline{R_{0}\big(\E^{-\I t\sqrt{u^2-m^2}}\,\, \ov{g(u)}\big)}$$
holds for the operator $\overline{R}_{0}$ given by
\begin{equation}
\overline{R}_{0}f=\frac{1}{2}
\frac{d}{d\ve}\int_{0}^{\infty}f(u)\overline{R_{0}(\ve,u)}du.
\label{5.39}\end{equation}
Thus, it follows from \eqref{5.35} and \eqref{5.36} that
\begin{align}& \overline{R}_{0}\big(\E^{\I t\sqrt{u^2-m^2}}g(u)\big){\sim} \I \frac{\pi\ve}{\lambda}|t|^{\I \phi(\lambda)}\E^{\I t\ve}\rho(\lambda)g(\lambda),
\quad t{\to}-\infty;
\label{5.37}
\\ &
\overline{R}_{0}\big(\E^{\I t\sqrt{u^2-m^2}}g(u)\big){\sim}0,
\quad t{\to}+\infty.\label{5.38}\end{align}

Step 3. Now, we return to the study of $T$. Relations \eqref{3.7},  \eqref{5.1}--\eqref{5.5} and  \eqref{5.35}--\eqref{5.38} imply that
\begin{equation}T\big(\E^{\I t\sqrt{u^2-m^2}}g\big){\sim}\frac{\pi\ve}{2\lambda}|t|^{-\I \phi(\lambda)}\E^{\I t\ve}\rho(\lambda)
\big(c_{2,1}(k,\lambda)\beta(\lambda)+\I c_{1,1}(k, \lambda)\big)g(\lambda),
\label{5.40}\end{equation}
where $t{\to}+\infty.$
According to \eqref{2.5}, \eqref{4.8} and \eqref{4.36} the equality
\begin{equation}c_{2,1}(k,\lambda)\beta(\lambda)+\I c_{1,1}(k,\lambda)=\I \frac{2\lambda}{\lambda+m}c_{1,1}(k,\lambda)
\label{5.41}\end{equation} 
is valid. Hence, relation \eqref{5.40} takes the form
\begin{equation}T\big(\E^{\I t\sqrt{u^2-m^2}}g\big){\sim}\I |t|^{-\I\phi(\lambda)}\E^{\I t\ve}
(\rho(\lambda)/\rho_{1}(\lambda))
c_{1,1}(k,\lambda)g(\lambda),
\label{5.42}\end{equation}
where  $t{\to}+\infty$ and the function $\rho_{1}(\lambda)$
is given by  \eqref{4.15}.
It follows from \eqref{5.37} and \eqref{5.41} that
\begin{equation}T\big(\E^{\I t\sqrt{u^2-m^2}}g\big){\sim}- \I |t|^{\I\phi(\lambda)}\E^{\I t\ve}
\big(\rho(\lambda)/\rho_{1}(\lambda)\big)
\overline{c_{1,1}(k,\lambda)}g(\lambda),
\label{5.43}\end{equation}where  $t{\to}-\infty.$
Using \eqref{5.42}, \eqref{5.43} and the invariance principle for generalized wave operators (see \cite[ Theorem 1.1]{Sakh4}),
we obtain the relations
\begin{align}& T\big(\E^{\I tu}g\big){\sim}\I |t\lambda/\ve|^{-\I \phi(\lambda)}\E^{\I t\lambda}
\big(\rho(\lambda)/\rho_{1}(\lambda)\big)c_{1,1}(k,\lambda)g(\lambda) \quad (t{\to}+\infty),
\label{5.44}
\\ &
T\big(\E^{\I tu}g\big){\sim}- \I |t\lambda/\ve|^{\I\phi(\lambda)}\E^{\I t\lambda}
\big(\rho(\lambda)/\rho_{1}(\lambda)\big)\overline{c_{1,1}(k,\lambda)}g(\lambda) \quad
(t{\to}-\infty).
\label{5.45}\end{align}

Step 4.
According to  \eqref{5.44} and \eqref{5.45} we have
\begin{align}& \lim_{t\to +\infty}\big(W_0(t)\E^{\I t \cll_0}\E^{-\I t \cll}\big) P =\I U_{0}
\big(\rho(\lambda)/\rho_{1}(\lambda)\big)c_{1,1}(k, \lambda)U^{-1}P,
\label{5.46}
\\ &
\lim_{t\to -\infty}\big(W_0(t)\E^{\I t \cll_0}\E^{-\I t \cll}\big) P=-\I U_{0}
\big(\rho(\lambda)/\rho_{1}(\lambda)\big)\overline{c_{1,1}(k, \lambda)}U^{-1}P,
\label{5.47}
\\ &
W_{0}(t)=|t\lambda / \ve|^{\I({\mathrm{sgn}}\, {t})\phi(\lambda)} .\label{5.53}
\end{align}
Definition \ref{Definition 1.1} of the generalized wave operators $W_{\pm}$,
relations \eqref{5.46}--\eqref{5.53} and the fact that $\cll_0$ and $\cll$ are self-adjoint,
 imply that
 \begin{align}& W_{+}(\mathcal{L},\mathcal{L}_{0}) =-\I U
\big(\rho(\lambda)/\rho_{1}(\lambda)\big)\ov{c_{1,1}(k, \lambda)}U_0^{-1}P_0,
\label{5.48}
\\ &
W_{-}(\mathcal{L},\mathcal{L}_{0})=\I U
\big(\rho(\lambda)/\rho_{1}(\lambda)\big)c_{1,1}(k, \lambda)U_0^{-1}P_0.
\label{5.48'}
\end{align}
Therefore, in view of \eqref{1.14}, the generalized dynamical scattering operator has the form
\begin{equation}S_{dyn}(\mathcal{L},\mathcal{L}_{0})=
U_{0}\big((\rho(\lambda)/\rho_{1}(\lambda))c_{1,1}(k,\lambda)\big)^{2}U_{0}^{-1}.
\label{5.49}\end{equation}
The scattering operator
$S_{dyn}(\mathcal{L},\mathcal{L}_{0})$ is unitary. Hence, it follows from
\eqref{5.49} that
\begin{equation}|c_{1,1}(k,\lambda)|=\rho_{1}(\lambda)/\rho(\lambda).\label{5.50}
\end{equation}Formulas \eqref{5.49} and \eqref{5.50} imply the following
representation of the generalized scattering operator
\begin{equation}S_{dyn}(\mathcal{L},\mathcal{L}_{0})=
U_{0}\big(c_{1,1}(k,\lambda)\big/\overline{c_{1,1}(k,\lambda)}\, \big)U_{0}^{-1},\quad \lambda>m.
\label{5.51}\end{equation} 
In the same way, it can be proved  that
\begin{equation}S_{dyn}(\mathcal{L},\mathcal{L}_{0})=
-U_{0}\big(c_{1,1}(k,\lambda)\big/\overline{c_{1,1}(k,\lambda)}\, \big)U_{0}^{-1},\quad \lambda<-m.
\label{5.52}\end{equation}
Recall that $S_{st}$ satisfies \eqref{3.8} and \eqref{4.38}. Then, formulas \eqref{d1},\eqref{d0} and 
\eqref{5.51}, \eqref{5.52} yield \eqref{d2}
 (i.e., the assertion of the theorem is proved).
\end{proof}
The proof of Theorem \ref{Theorem 5.1} and Definition \ref{Definition 1.1} imply the assertion.
\begin{Cy}\label {Corollary 5.2} Let the conditions of Theorem 5.1 hold. Then,
the  deviation factor $W_{0}(t)$ corresponding to the operators $\cll$ and $\cll_0$ has the form
\eqref{5.53}.

\end{Cy}
Comparing equalities \eqref{4.35} and \eqref{5.53}, we obtain the assertion:
\begin{Cy}\label{Corollary 5.3} The deviation factor $V_{0}(r,\la)$
for the stationary case and  the deviation factor $W_{0}(t)=W_0(t,\la)$
for the dynamical  case are connected by the following simple equality:
\begin{equation}V_{0}(|t \lambda/\ve|,\la)=W_{0}(t,\la),\quad t>0.\label{5.54}\end{equation}\end{Cy}
\section{The classical case ($A=0$)}\label{secA}
In this section, we again consider  the operator $\mathcal{L}$ of the form \eqref{4.25}, \eqref{4.26}, where $v$ in \eqref{4.26} is given by \eqref{2.3}, but
this time we set $A=0$ in \eqref{2.3}.   That is, we consider the classical case. Dynamical and stationary approaches for this case were studied
separately in many important publications (see, e.g., \cite{BiKr, Kato, Rose}).
Here, we  compare  these approaches, and our ergodic-type theorem is new even for the classical case. We stress that  the classical wave and scattering operators are used
in this section instead of the generalized wave and scattering operators in Section \ref{ErgProp}. The result and proof are similar
to Theorem \ref{Theorem 5.1} and its proof but there are some differences, and so we consider the case $A=0$ separately.
\begin{Tm}\label{Theorem 6.1} Let the radial Dirac system \eqref{2.1}, \eqref{2.2}, where $v(r)\equiv q(r)$ and \eqref{2.23} holds,
be given. 

Then $($for the
corresponding operators $\cll_0$ and $\cll$ defined via \eqref{4.16}, \eqref{4.18} and  \eqref{4.25}--\eqref{4.27}, respectively$)$ 
the generalized stationary and dynamical scattering
matrices are equal on $L$, that is, 
$
S_{st}(\mathcal{L},\mathcal{L}_{0})f=S_{dyn}(\mathcal{L},\mathcal{L}_{0})f$ for $f\in L$, where $S_{st}$, $S_{dyn}$ and $L$
are given in \eqref{4.38}, \eqref{d1} and \eqref{d0}, respectively.
\end{Tm}
\begin{proof}. First, we consider the case \eqref{2.16} where $\la >m$. Recall that $\ve$ is determined in \eqref{2.5}. According to \cite[Ch.8, f-las
495 and 496]{BrP}, the following Fourier transformation equalities are valid: 
\begin{align}& \int_{-\infty}^{+\infty}\E^{-\I r\sqrt{u^2-m^2}}\frac{(\sin(\ve r/2))^2}{r}dr=
\frac{\pi}{2\I}\begin{cases}
1, &\text{$m<u<\lambda$;}\\
0, &\text{$u>\lambda$.}
\end{cases}
\label{6.2}
\\ &
\int_{-\infty}^{+\infty}\E^{-\I r\sqrt{u^2-m^2}}\frac{(\sin(\ve r/2))^2}{|r|}dr=
\frac{1}{2}\ln\left|\frac{u^2-\lambda^{2}}{u^2-m^2}\right|.
\label{6.3}\end{align}
From \eqref{6.2} and \eqref{6.3}, we derive
\begin{equation}\int_{0}^{+\infty}\E^{-\I r\sqrt{u^2-m^2}}\frac{(\sin(\ve r/2))^2}{r}dr=
\frac{1}{4}\ln\left|\frac{u^2-\lambda^{2}}{u^2-m^2}\right|-
\frac{\I\pi}{4}\begin{cases}
1, &\text{$m<u<\lambda$;}\\
0, &\text{$u>\lambda$.}
\end{cases}
\label{6.4}\end{equation}
Following the scheme of the proof of Theorem \ref{Theorem 5.1}, we rewrite  \eqref{5.5}  in the form
\begin{equation}T_{2}(\ve,u)=
\rho(u)\int_{0}^{\infty}
g(r,u)\frac{2(\sin(r\ve/2))^2}{r}dr.
\label{6.5}\end{equation}
In view of \eqref{6.2}--\eqref{6.4} the corresponding operator $R_{0}$
is defined by the relation
\begin{equation}R_{0}g=\frac{d}{d\ve}\int_{0}^{\infty}g(u)
\Phi(\ve,u)du,\label{6.6}\end{equation}
where
\begin{equation}\Phi(\ve,u)=
\frac{1}{2}\rho(u)\ln\left|\frac{u^2-\lambda^{2}}{u^2-m^2}\right|-
\frac{\I \pi}{2}\begin{cases}
\rho(u), &\text{$m<u<\lambda$;}\\
0, &\text{$u>\lambda$.}
\end{cases}
\label{6.7}\end{equation}
 We represent the operator $R_0$ in the form
 \begin{equation}R_0=R_{-}+R_{+},\label{6.8}\end{equation}
 where
\begin{align}& R_{-}g=\frac{1}{2}\frac{d}{d\ve}\left(-\I \pi\int_{m}^{\lambda}g(u)\rho(u)du
+\int_{m}^{\infty}g(u)\rho(u)\ln\left|u-\lambda\right| du\right),\label{6.9}
\\ &
R_{+}g=\frac{1}{2}\frac{d}{d\ve}\int_{m}^{\infty}g(u)
\rho(u)\left(\ln(u+\lambda)-\ln(u^2-m^2)\right)du .\label{6.10}\end{align}
It is easy to see that relation \eqref{5.17} is valid in the case $A=0$ too.
Taking into account \eqref{6.9} and equality $\lambda=\sqrt{\ve^2+m^2}$,
we obtain
\begin{equation}R_{-}g=\frac{\ve}{2\lambda}\left(-\I \pi\rho(\lambda)g(\lambda)+
\fint_{m}^{\infty}\frac{g(u)\rho(u)}{\lambda-u}du\right).\label{6.11}\end{equation}
We note that the integral $\fint$  on the right-hand side of  \eqref{6.11}
is a Cauchy-type integral.
Using \eqref{6.8}, \eqref{6.10}, \eqref{6.11} and the equality
\begin{equation}\lim_{t{\to}\pm\infty}\frac{\I}{\pi}\fint_{m}^{\infty}f(u)
\frac{\E^{\I(\lambda-u)t}}{\lambda-u}du={\mp}f(\lambda),\label{6.12}\end{equation}
one may show that the relations
\begin{equation}R\left(\E^{\I t\sqrt{u^2-m^2}}g\right){\sim}-2\I{\pi}\E^{\I t\ve}\rho(\lambda)g(\lambda),
\quad t{\to} - \infty;
\label{6.13}\end{equation}
and
\begin{equation}R\left(\E^{\I t\sqrt{u^2-m^2}}g(u)\right){\sim}0,
\quad t{\to} + \infty
\label{6.14}\end{equation}
are valid.
The final part of the proof of Theorem \ref{Theorem 6.1} coincides with  the final part of the proof of Theorem \ref{Theorem 5.1}.
\end{proof}

\section{Appendix}  \label{Appendix }
\emph{In this Appendix, we introduce the notions of the generalized dynamical wave and scattering operators}
\cite{Sakh8, Sakh2, Sakh4} (see also \cite{BuM}).
Consider linear (not necessarily bounded)
operators $A$ and $A_{0}$ acting in some Hilbert space $H$  and assume that the operator $A_{0}$
is self-adjoint.  The absolutely continuous subspace of the operator $A_0$ (i.e., the subspace corresponding
to the absolutely continuous spectrum) is denoted by $G_0$, and $P_0$ is the orthogonal projection on   $G_0$.
Generalized wave operators $W_{+}(A,A_{0})$ and $W_{-}(A,A_{0})$ are introduced by the equality
\begin{equation}W_{\pm}(A,A_{0})
=\lim_{t{\to}\pm\infty}\big(\E^{\I At}\E^{-\I A_{0}t}W_{0}(t)^{-1}\big)P_0,
\label{1.1a}\end{equation}
where  $W_0$ is an operator function taking operator values $W_0(t)$   acting in $G_0$
in the domain $|t|>R$ ($t \in \BR$) for  some $R \geq 0$.  
More precisely, we have the following definition (see \cite{Sakh2, Sakh8}) of
 the generalized wave operators $W_{\pm}(A,A_{0})$ and deviation factor $W_0$.
 \begin{Dn}\label{Definition 1.1}
An operator function $W_{0}(t)$ is called a deviation factor 
and operators $W_{\pm}(A,A_{0})$ are called generalized wave operators if
the
following conditions are fulfilled:
\begin{enumerate}
	\item The operators $W_{0}(t)$ and  $W_{0}(t)^{-1}$  acting in  $G_0$,
are
bounded for all $t$ \, $(|t|>R)$, and
\begin{equation}\lim_{t{\to}\pm\infty}W_{0}(t+\tau)W_{0}(t)^{-1}P_0=P_0,\quad\tau=\overline{\tau}.
\label{1.2}
\end{equation}
\item The following commutation relations  hold for arbitrary values $t$ and $\tau$:
\begin{equation}W_{0}(t)A_{0}P_0=A_{0}W_{0}(t)P_0,\quad
W_{0}(t)W_{0}(t+\tau)P_0=W_{0}(t+\tau)W_{0}(t)P_0.
\label{1.3}\end{equation}
\item The limits $W_{\pm}(A,A_{0})$ in \eqref{1.1a}
exist in the sense of strong convergence. 
\end{enumerate}
\end{Dn}
If  $W_{0}(t) \equiv I$ in $G_0$ (where $I$ is the identity operator), then the operators $W_{\pm}(A,A_{0})$ are usual wave
operators.

Clearly, the choice of the deviation factor is not unique.

\begin{Rk}\label{Remark 1.8} 
Let unitary operators $C_{-}$ and $C_{+}$ satisfy commutation conditions $A_{0}C_{\pm}=C_{\pm}A_{0}$. If $W_{0}(t)$ is a deviation factor, then
the operator function given $($for $t>0$ and $t<0$, respectively$)$ by the equalities
$W_{+}(t)=C_{+}W_{0}(t)$\, $(t>0)$), and $W_{-}(t)=C_{-}W_{0}(t)$\, $(t<0)$) is the deviation factor as well.
\end{Rk}
The choice of the operators $C_{\pm}$ is very important and is determined by specific physical problems.
The definition below shows that generalized scattering operators also depend on the choice of $C_{\pm}$.
\begin{Dn}\label{Definition 1.9}
The generalized scattering operator $S(A,A_0)$ has the form
\begin{equation}S(A,A_0)=W_{+}(A,A_0)^{*}W_{-}(A,A_0),\label{1.14}
\end{equation}
where
\begin{equation}\nn W_{\pm}(A,A_{0})
=\lim_{t{\to}\pm\infty}\Big( \E^{\I At}\E^{-\I A_{0}t}W_{\pm}(t)^{-1}\Big) P_0 .
\end{equation}
\end{Dn}

In fact, operator functions $W_{\pm}(t)$ are uniquely determined up to some factors $C_{\pm}(t)$
tending to $C_{\pm}$ when $t$ tends to $\infty$ or $- \infty$, respectively. This means that
 $S(A,A_0)$ is uniquely determined by the choice of $C_{\pm}$.
 
It is not difficult to prove that the operator $S(A,A_0)$ unitarily  maps $G_0$  onto itself and that
\begin{equation}
A_{0}S(A,A_0)P_{0}=S(A,A_0)A_{0}P_{0}.
\label{1.16}
\end{equation}

\vspace{0.3em}

{\bf Acknowledgements.}
The author is grateful to  A. Sakhnovich and I.~Roitberg for fruitful discussions
and help in the preparation of the manuscript.

\end{document}